\documentstyle[sprocl,psfig]{article}

\def\Journal#1#2#3#4{{#1} {\bf #2}, #3 (#4)}

\def\PRL{\em Phys. Rev. Lett.}
\def\PRB{{\em Phys. Rev.} B}

\def\beq{\begin{equation}}
\def\eeq{\end{equation}}
\def\bea{\begin{eqnarray}}
\def\eea{\end{eqnarray}}
\def\nnu{\nonumber}

\def\be{\beta}

\def\sig{\sigma}

\def\tta{\theta}
\def\om{\omega}
\def\ptl{\partial}

\def\adag{a^{\dagger}}
\def\bdag{b^{\dagger}}

\def\ham{{\cal H}}
\def\hhf{\ham_{\rm hf}}
\def\Dta{\Delta}
\def\zhat{\hat{\bf z}}
\def\xhat{\hat{\bf x}}

\def\half{{1\over 2}}
\def\tofro{\leftrightarrow}
\def\sech{{\rm sech\,}}
\def\bS{{\bf S}}
\def\bfI{{\bf I}}
\def\ket#1{|#1\rangle}
\def\bra#1{\langle#1|}
\def\tran#1#2{\langle#1|#2\rangle}
\def\avg#1{\langle#1\rangle}
\def\mel#1#2#3{\langle#1|#2|#3\rangle}

\def\kI{\ket{{\rm I}}}
\def\kII{\ket{{\rm I}{\rm I}}}
\def\kIII{\ket{{\rm I}{\rm I}{\rm I}}}
\def\kIV{\ket{{\rm I}{\rm V}}}
\def\bI{\bra{{\rm I}}}
\def\bII{\bra{{\rm I}{\rm I}}}
\def\bIII{\bra{{\rm I}{\rm I}{\rm I}}}
\def\bIV{\bra{{\rm I}{\rm V}}}

\def\xx{\times}
\def\efrm#1#2{#1\xx 10^{-#2}}
\def\efrt#1#2#3{#1\xx 10^{-#2}(#3)}

\begin{document}

\title{COHERENT MAGNETIC MOMENT REVERSAL IN SMALL PARTICLES
WITH NUCLEAR SPINS}

\author{ANUPAM GARG}

\address{Department of Physics and Astronomy \\ Northwestern University\\
2145 Sheridan Road\\ Evanston, IL 60208, USA\\E-mail: agarg@nwu.edu} 

\maketitle\abstracts{A discrete WKB method is developed for calculating
tunnel splittings in spin problems. The method is then applied to the
issue of how nuclear spins affect the macroscopic quantum coherence of
the total magnetic moment in small magnetic particles. The results are
compared with numerical work, and with previous instanton based analytic
approaches.}

\section{Motivation for this Work}
Small magnetic particles have now been investigated as good candidates for
experimental observation of macroscopic quantum tunneling (MQT) and coherence
(MQC).~\cite{cg,ajl,ag1} This is because at first sight the main critera for
a system to be a good candidate for seeing MQP (P for phenomena) are met.
One of these is that the energy barrier through which the system must tunnel
be microscopic, even though the tunneling variable itself is macroscopic.
This demand is met for the physical reason that the anisotropy energy barrier
originates in spin-orbit or spin-spin interactions at the
microscopic level, both of which are relativistic effects, and hence small.
A second criterion is that there be a well defined macrovariable, whose
dynamics can, to a first approximation,
be isolated from those of other microscopic degrees of freedom.
This criterion can be met by working at temperatures sufficiently below the
equivalent anisotropy energy gap, as spin wave excitations are then frozen
out, and we may focus on the net magnetic moment of the particle as the
macrovariable.

There are also extremely good reasons to believe, however, that MQP are
very hard to observe in general. Chief among these reasons is that the couplings
of the macrovariable to the microscopic degrees of freedom give rise to
decoherence. This effect can be especially severe in the case of macroscopic
quantum coherence, even when the coupling is so
small that the semiclassical dynamics of the macrovarible are significantly
underdamped. The best studied example which provides a detailed illustration
of this point is that of the spin-boson problem.~\cite{chl,smr,uw}
Further support for this point is provided by the {\em scantiness of observational
evidence} for quantum coherence even in systems that one would regard as
microscopic. Thus, we know of only a few dozen or so ``flexible'' molecules
like NH$_3$ which display coherent flip-flop between different nuclear
configurations.~\cite{hwk} The frequency of flip-flop is generally in the
0.1--100~GHz range. Given the vast range of molecular structures, and bonding
strengths, there must surely be many naturally occurring molecules whose energy
barriers and attempt frequencies are such as to put the flip-flop frerquency at
anout 1~Hz. Yet such flip-flop has never been seen. The reason almost certainly
is that such molecules are never naturally encountered in isolation by themselves,
and collisions and other environmental interactions are very effective in wiping
out the quantum coherence.

It is thus important to identify mechanisms for decoherence in the magnetic
particle system, and several have been put forth (phonons, magnons, Stoner
excitations). The most critical, however, is the spin of the nuclei in the
particle. The hyperfine coupling between the nuclear and electronic spins
in magnetic solids is of order 100~MHz or more (in frequency units) per nucleus,
which is rather high on the scale of the expected MQC frequencies. At the same time
this frequency is rather low on the scale of the attempt frequencies associated
with the electronic moments. Nuclear spins are therefore likely to be extremely
efficient decoherers or ``observers'' of the direction of the moment of a small
particle. This expectation is confirmed by theoretical calculations for both
MQT,~\cite{ag2} and MQC.~\cite{ag3} It is the latter that I wish to focus on
and revisit in this article, for several reasons. First is that this decoherence
mechanism falls outside the scope of the harmonic oscillator bath,~\cite{fv,cl}
so one cannot rely on previous results. The calculations in
Ref.~9 are done using an instanton technique with many physically
motivated approximations about the trajectories likely to give the dominant
contribution to some path integral. It is not obvious even to me
that this calculation is
done in strict adherence with the {\em Rheinheitsgebot}.
On the occasions that I have given seminars
on the subject, the quizzical looks on the faces of my audience make it clear
to me that it does not fully believe or understand my approach. It is
therefore desirable to study this problem using different methods, and we
shall do so in this article using the discrete WKB method.
Although this method itself is quite old (see Braun's review~\cite{pab} for
references), it does not appear to have been used in spin tunneling problems
except for some work by van Hemmen and S\"ut\H o.~\cite{vHS} These authors have
not fully exploited the power of this method, however.  We shall see that
calculations which have traditionally been done by instanton methods can be done
much more simply to the same accuracy using this method. The opportunity
to present the technical aspects of this method in a tutorial
volume devoted to tunneling in complex systems is greatly welcome, and
provides me with another reason for writing this article.

It should be noted that the same subject has also been studied by Prokof'ev and
Stamp in several papers. The first of these~\cite{ps1} correctly notes that
the tunneling amplitude is suppressed by nuclear spins, but fails to recognize
the implications of this fact for the nature of the tunneling spectrum. See
footnote {\em d} for more on this.
Subsequent papers~\cite{psmany} puport to make detailed calculations
of the tunneling spectrum including lineshapes, and also to include a host of
other physical effects, such as spin diffusion, the Suhl-Nakamura interaction, and
others with less familiar names such as ``topological decoherence",
``orthogonality blocking", and ``degeneracy blocking''. I cannot comment on the later
work, simply because I do not understand much of it, especially some of the more
mathematically specific and detailed conclusions, about
lineshapes, for example. My goal
in this article will be much more modest.  It was argued in Refs. 3 and 9 that
in the presence of nuclear spins the tunneling spectrum is broken into several
resonance lines. A physical interpretation was attached to this broken-up spectrum,
and formulas were presented for their frequencies and spectral weights. 
This article corroborates these claims. As in the previous papers, I have not
attempted to model or calculate the details of the relaxation, i.e., the lineshapes.
To this extent, I am only prepared to claim a qualitative
understanding for the very low frequency part of the tunneling spectrum. Fortunately, it
is the high frequency end that is most likely to be experimentally relevant if at
all, and here the situation is much better. 

The plan of this article is as follows. In Sec.~2, I will give a brief introduction
to the physical problem and the models studied in this paper. Sec.~3 contains a general
discussion of the discrete WKB method. This is used to calculate the bare tunnel
splitting, i.e., without nuclear spins, in Sec.~4. Nuclear spins are added to the
problem in Sec.~5. This section has the bulk of the new results in this article.
The results for the tunnel splitting(s) obtained via the discrete WKB method are
checked against those from exact numerical diagonalization of model Hamiltonians.
I also compare them with those from the instanton approach,~\cite{ag3}
and identify which features of the latter appear
to be quantitatively robust, and which are only qualitatively correct. I conclude
in Sec.~6 with a summary of the effects of nuclear spins, and some general remarks
on the observability of MQC in magnetic particles.

\section{Introduction to Physical Problem and Models}
\subsection{Physical System}
The physical system is a small insulating magnetic particle, about $50\ $\AA\ in diameter,
at millikelvin temperatures. The anisotropy energy gap is much larger than this, so
spin wave excitations are frozen out, and the individual atomic spins are
orientationally locked together. Furthermore, at this size the particle
typically contains only one magnetic domain. Thus the magnitude of the total spin or
magnetic moment of the particle is essentially fixed proportional to the number
of atomic spins (assuming one magnetic species for simplicity), and the only relevant
dynamical variable is its direction. (It may be useful to think of the system 
in terms of a Heisenberg-like model, with additional
single-ion anisotropy terms.) The existence
of anisotropy implies that some directions are energetically favored over others, 
and we wish to investigate whether the spin orientation can display quantum mechanical
behavior, in particular tunneling and/or coherent oscillation between different
energy minima.

A model Hamiltonian which incorporates the above features is:~\cite{cg}
\beq
\ham_0 = -k'_1 S_z^2 + k'_2 S_x^2
\label{ham0}
\eeq
Here $S_x$, $S_y$, and $S_z$ are the components of our large spin, and $k'_i$
are phenomenological anisotropy coefficents. This particular Hamiltonian is time-reversal
invariant, which should be the case if our particle is not subject to any external
magnetic fields. We take $k'_1>0$, $k'_2 > 0$, so that in classical language, $\pm\zhat$
are easy directions, and $\pm\xhat$ are hard directions. Quantum mechanically, the
two classical ground states $\pm\zhat$ will be split by tunneling. A generally
valid approximate expression for the tunnel splitting can be obtained by writing
\beq
\Dta_0 \simeq \om_e \exp(-S u/\om_e).
\label{spapr}
\eeq
[A more exact expression is given in Eq.~(\ref{Dtaex}) below.] In this formula,
\beq
\om_e = 2(k_1k_{12})^{1/2}
\label{fel}
\eeq
is the oscillation (or precession) frequency for small deviations of the spin
orientation from the classical equilibrium directions
 $\pm\zhat$,\footnote{This result may be derived
by writing down the Heisenberg equation of motion for $\bf S$, and linearizing it in
small deviations from ${\bf S} = S\zhat$.} and we have also defined
\beq
k_1 = S k'_1, \quad k_2 = S k'_2, \quad k_{12} = k_1 + k_2.
\label{kint}
\eeq
Further, $Su$ is a quantity proportional to the energy barrier, of order $S k_{12}$. 
We work throughout in units such that $\hbar =1$.

We next wish to consider the influence of nuclear spins. To model this
let us suppose that $N_n$ of
the atoms have nuclei with spins ${\bf I}_i$, and take all of these to be
of magnitude $1/2$. Let us further simplify the problem and assume that the hyperfine
interaction for each of these atoms is of identical strength and of the form
${\bf s}_i\cdot{\bf I}_i$, where ${\bf s}_i$ is the electronic spin on atom $i$.
All the assumptions except that of identical interaction strength are immaterial,
and even this is a rather good approximation. The effects of relaxing it will be
briefly discussed in Sec.~5.3.
The total hyperfine contribution to the Hamiltonian can then be written as
\beq
\hhf = -{\om_n \over s}\sum_{i=1}^{N_n} {\bf s}_i\cdot{\bf I}_i
          = -{\om_n\over S}{\bf S}\cdot {\bf I}_{\rm tot},
\label{hhf1}
\eeq
where we have expressed  the coupling constant in terms of $\om_n$, the nuclear Larmor
frequency that would be obtained if the electronic spin orientation were fixed.
The first expression in Eq.~(\ref{hhf1}) is just the sum of the interactions
for the individual atoms, and the second
follows from assuming that the atomic spins are all parallel to one another, which
permits one to write ${\bf s}_i = (s/S){\bf S}$ for all $i$. The remaining sum equals
$\sum_i {\bf I}_i$, which we call ${\bf I}_{\rm tot}$, the total nuclear spin. For any
given value of $N_n$, $I_{\rm tot}$ can take on values ranging
in integer steps from $N_n/2$ to either 0 (if $N_n$ is even) or $1/2$ 
(if $N_n$ is odd), with multiplicities that can easily be found. (See below.)
Since $\hhf$ commutes with ${\bf I}^2_{\rm tot}$, we can consider the
problem for one value of $I_{\rm tot}$ at a time. Writing ${\bf I}$ instead of
${\bf I}_{\rm tot}$ henceforth, we arrive at a model Hamiltonian obtained by
adding Eqs.~(\ref{ham0}) and (\ref{hhf1}), i.e.,
\beq
\ham_I = {1 \over S}\left(-k_1 S_z^2 + k_2 S_x^2
           -\om_n {\bf S}\cdot {\bf I}\right).
\label{hamI}
\eeq

[For completeness, we give here the formula for the multiplicity, i.e.,
the number of times a given value of $I_{\rm tot}$ appears when $N_n$
spins of magnitude 1/2 are added together.
For $I_{\rm tot} = (N_n/2) - k$, the multiplicity is given by
\beq
{N_n \choose k} - {N_n \choose k-1}. \label{multi}
\eeq
As an example,
if $N_n =6$, multiplets with $I_{\rm tot} = 3$, 2, 1, and 0, occur 1, 5, 9, and 5
times respectively.] 

It should be noted at this point that for a particle of diameter $50\ $\AA, which
corresponds to $S={\cal O}(10^4)$, and typical material or anisotropy parameters,
the tunnel splitting $\Dta_0$ as given by Eq.~(\ref{spapr})
is unobservably small. Of course the splitting goes
up if $S$ is decreased, but particles much smaller than $50\ $\AA\ seem difficult to
attain controllably and reproducibly with present day technology.\footnote{The last
few years have seen some very interesting work\cite{mn1,mn2,mn3} on systems with much
smaller values of $S$, of order 10. These systems are based on magnetic molecules,
where the value of $S$ is highly reproducible, but it is my belief that the
interesting questions here are quite different, and have little to do with MQP.
Interestingly, semicalssical methods, such as those developed in this paper, are
likely to be quite valuable in analyzing these systems. For particles with intermediate
values of $S$, say about 100, on the other hand, it will probably be necessary to
devise new ways of treating the environment because the law of
large numbers will no longer be applicable.}
MQC is a more likely proposition in {\em antiferromagnetic}
particles\cite{bc,kz,dg} where the two states involved differ in the orientation
of the N\'eel vector, or the spins on one of the sublattices.\footnote{The reason
for this is quite simple. The tunnel splitting can still be written in the form
(\ref{spapr}). The energy barrier $Su$ is of the same order of magnitude,
 but the electronic spin attempt frequency, $\om_e$,
is about 100 times higher for antiferromagnets than ferromagnets, being given by
the geometric mean of an exchange energy and an anisotropy energy in contrast
to Eq.~(\ref{fel}).} The essential aspects of the earlier papers,~\cite{ag1,ag3}
in which the influence of nuclear spins was studied using instantons, hold
equally for ferro- and antiferro-magnetic particles.
My goal in this article is to try and verify the predictions of the instanton
approach as quantitatively as possible. It is simpler to do this using a ferromagnetic
model, and I shall therefore limit myself to that in this article.   

I also do not wish to discuss at length the experimental observability of the reversal
phenomenon. In addition to the values of the physical parameters mentioned above,
this hinges on a number of other issues such as the temperature, the
signal size, stray magnetic fields, the nature of the substrate, etc.
We refer readers to previous
papers~\cite{ag1,ag3,ag4,ag5} for detailed discussions of these points. For
the purposes of this article, we will only note that the
ratio $\om_n/\om_e$ is of order $10^{-3}$ to $10^{-2}$ in antiferomagnets, and
is about one order of magnitude higher in ferromagnets. Throughout this
article therefore, we shall assume that $\om_n/\om_e \ll 1$, and work to leading
order in this ratio. We shall further assume that $\Dta_0/\om_n \ll 1$, which,
by virtue of the exponentially small WKB or Gamow factor, is
almost certain to be the case for values of $S$ of interest to us.

\subsection{What to Calculate; Preliminary Arguments}
Having settled on a theoretical model,
let us ask what physical quantity we should calculate. One way the
dynamical behaviour of the moment can be described is in terms of a time dependent
probability to find it along some direction given an initial state in which it
was prepared.\cite{chl,smr} This description best applies to
experiments on a single system. Another way is to find
the appropriate frequency-dependent dynamical
susceptibility $\chi(\om)$. This description
better applies to an assembly of identical or nearly identical systems,
such as is obtained in an NMR experiment. We will adopt the second approach.

Let us suppose then that we wish to calculate $\chi''(\om)$ for our magnetic
particle. Since the Hamiltonian (\ref{hamI}) describes a closed system, $\chi''$
can only consist of a set of delta functions. The real system is of course not
closed, and has means of energy relaxation such as phonons, with which the spins can
interact in a variety of ways. These will lead to line widths and broadening in
the usual way. Now in a setup such as NMR, the resonance frequencies
are few in number and usually known quite accurately to begin with. Interest then
attaches to the calculation of line shifts and widths and shapes and so on. In our
problem, on the other hand, the presence of nuclear spins alters
the resonance frequencies themselves drastically. To see this, let us denote by
$\ket{\pm 0}$ the approximate eigenstates of the bare Hamiltonian $\ham_0$ that
correspond to the spin having a mean value $\avg{S_z} \approx \pm S$, i.e., to
states in which the spin is localized in the ground state in one of the two energy
minima centerd at $\pm\zhat$.
Tunneling mixes these states and splits them by $\Dta_0$. In addition
to these states, let us consider the next few excited states in each well, and denote
them by $\ket{\pm 1}$, $\ket{\pm 2}$, and so on. Since the small oscillations (or more
accurately precessions) around the $\pm\zhat$ directions are approximately
harmonic, these states will form an approximately equally
spaced ladder with spacing $\om_e$, i.e., the states $\ket{\pm n}$ will lie an energy
$\om_e$ above $\ket{\pm(n-1)}$ for low values of $n$. (See Fig.~1.)
\begin{figure}
\centerline{\psfig{figure=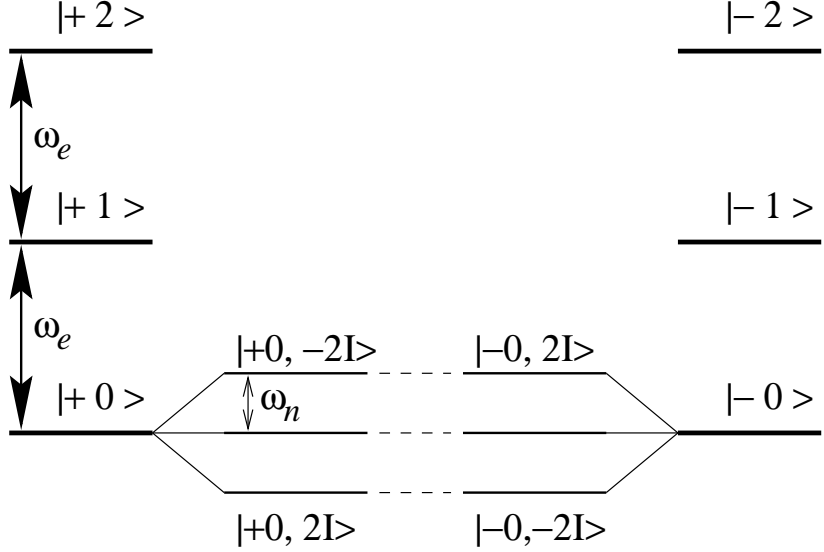}}
\caption{Structure of low lying energy levels of coupled electron and nuclear spin
systems. The states $\ket{\pm 0}$ are electronic spin ground states with $\bS$
localized along $\pm\zhat$ in the absence of the nuclear spins. The corresponding
first two excited states are $\ket{\pm 1}$ and $\ket{\pm 2}$.
The central part of the figure shows how
the ground electronic levels are modified by the nuclear spins, still ignoring
tunneling. The ratio $\om_e/\om_n \gg 1$ and is not accurately represented in the figure.
If $\om_n \gg \Dta_0$, only states joined by dashed lines mix with each
other to any appreciable extent once tunneling is turned on.}
\end{figure}
Now let us bring in
the nuclear spins. Since $\bra{\pm 0}{\bf S}\ket{\pm 0} = \pm S\zhat$ to very good
accuracy, the combined elctronic-nuclear spin states will be eigenstates of
$I_z$, which we label by $\ket{+0,p}$,
and $\ket{-0,p}$, where $p=2I_z$, will be split into $2I+1$ Zeeman levels with a
spacing $\om_n$. (The higher states $\ket{\pm 1}$, $\ket{\pm 2}, \ldots$, will also be
similarly split, but we do not consider that for the moment.) As long as $I$ is
small enough so that $2I\om_n \ll \om_e$, our basic assumption that $\om_n\gg\Dta_0$
combined with the physically obvious but important fact that significant resonance
is only possible between states that are degnerate to within the matrix element
connecting them implies that the state $\ket{+0,p}$ will resonantly tunnel only
to the state $\ket{-0,-p}$. Since this tunneling now involves a change in the nuclear
spin state in addition to that of the electronic spin, the splitting should in general
be different from $\Dta_0$. We denote the magnitude of this
splitting by $\Dta(I,p)$, i.e.,
\beq
\Dta(I,p)= \pm (E_{g(0,p)} - E_{u(0,p)}),
\label{DtaIi}
\eeq
where $E_{g(0,p)}$ and $E_{u(0,p)}$ denote the energies of the antisymmetric
and symmetric linear combinations $(\ket{0,p} \mp \ket{0,p})/\sqrt2$. Note that the
antisymmetric state need not be the one with higher energy.

To orient further discussion, let us summarize the results of the instanton
approach.~\cite{ag1,ag3}  In this approach too the starting point is that the
nuclear spins spoil the degeneracy between the $\ket{\pm 0}$ states. An
instanton connecting degenerate states must involve the flipping over of
a certain number of nuclear spins along with that of $\bS$. Since the nuclear
spins can only respond on a time scale $\om_n^{-1}$ to any perturbation, and the
instanton has a temporal width of order $\om^{-1}_e$, perturbation
theory shows that each nuclear spin coflip reduces the tunneling amplitude by a factor 
$\sim (\om_n/\om_e)$.\footnote{This
reduction of tunneling amplitudes due to nuclear spin coflips was also found by
Prokof'ev and Stamp.~\cite{ps1} The highly chopped-up nature of the $\chi''$ spectrum
was missed by them, however, and the mechanism for decoherence initially
studied by them is rather different from mine.\cite{ag1,ag3}}
The amplitude for $p$ nuclear spin coflips was
found~\cite{ag1,ag3} (in the case of an antiferromagnetic particle) to be
\beq
\Dta_p =  (\pi\om_n/2\om_e)^{|p|} \Dta_0.
\label{Delpin}
\eeq
In other words, the tunneling amplitude decreases geometrically with the number
of units by which $I_z$ must change. If this is so (and we will find here that by
and large it is), then clearly the more interesting problem is find the tunneling
frequencies $\Dta(I,p)$
themselves, and the associated spectral weight to be assigned to each frequency. The
issue of linewidths and relaxation becomes secondary. As stated in the previous section,
however, the calculation on which these conclusions are based is not totally
satisfactory. It is with this viewpoint that we focus in this article on calculating
the splittings as carefully as possible.

It should be clarified that the amplitudes $\Dta_p$ are not to be identified with the
tunnel splittings themselves. To see this, consider a multiinstanton trajectory
that starts and ends at a specific state with $S_z = S$ and $I_z = p/2$. The intermediate
states with a long residence time in this trajectory are those with $S_z=S$, $I_z=p/2$,
and $S_z = -S$, $I_z = -p/2$. Since $p$ spins can be chosen from $N_n$ in many ways, 
there can be several such intermediate states, and the amplitudes for all the
corresponding trajectories must be added together. This leads to interference effects.
When the combinatoric factors are added together, it is found that the {\em splittings}
are given by~\cite{ag3}
\beq
{n\choose |p|}\Dta_p,\quad n=|p|,|p|+1,\ldots,(N_n+|p|)/2.
\label{splbi}
\eeq

In this article we will try and see how correct these results are. The main flaw in
the previous work is the conclusion that $\Dta_p = \Dta_{-p}$, i.e., that the
$\ket{0,p}\tofro \ket{-0,-p}$ and $\ket{0,-p}\tofro \ket{-0,p}$ splittings are
the same. In yet more words, the
splittings for $p$ coflips are independent of whether the coflipping spins are
parallel or antiparallel to the large electronic spin $\bS$. We will see that this
is no longer true. Further, the geometric dependence $(\om_n/\om_e)^{|p|}$ seems to
be only approximately correct. Finally, we will show in Sec.~5 that the
binomial factor in Eq.~(\ref{splbi}) follows from a simple effective tunneling
Hamiltonian connecting the states $\ket{0,p}$ and $\ket{-0,-p}$, which holds as
long as $\om_n/\om_e \ll 1$.

One trivial point should be noted and disposed of once and for all. Since the
Hamiltonian (\ref{hamI}) is time-reversal invariant, the bare tunnel splitting 
vanishes unless $S$ is an integer, and all the splittigns $\Dta(I,p)$
vanish unless $S+I$ is an integer. For the purposes of interpreting the
relation (\ref{Delpin}) (or a more accurate replacement) it is useful to
{\em define} $\Dta_0$ for all $S$ by Eq.~(\ref{spapr}) or its more
accurate version, Eq.~(\ref{Dtaex}) below.

\section{The Discrete WKB Method}
Like its continuous counterpart, the discrete WKB method is applicable to a wide
variety of settings. (See the review by Braun~\cite{pab} for examples.) We will only
give an overview and physically motivated discussion of this method using the
Hamiltonian (\ref{ham0}) as an illustrative example.

Let us write a general eigenfunction of $\ham_0$ as $\sum_m a_m\ket m$, where
as usual $S_z\ket m = m\ket m$. Schr\"odinger's equation then becoems a three-term
recursion relation for the coefficients $a_m$:
\beq
w_m a_m + t_{m,m+2}\,a_{m+2} + t_{m,m-2}\,a_{m-2} = E a_m,
\label{ttr}
\eeq
where $w_m = \mel m{\ham_0}m$, and $t_{m,m\pm 2} = \mel m{\ham_0}{m\pm 2}$.
Now for large $S$, the differences $w_{m+2} - w_m$ and $t_{m,m+2} - t_{m-2,m}$,
etc., are of order $1/S$ relative to $w_m$ and $t_{m,m+2}$ themselves. It is
thus extremely useful to view Eq.~(\ref{ttr}) as arising from a tight-binding
model for an electron on a one-dimensional lattice with sites labeled by $m$, 
on-site energies $w_m$, and hopping energies $t_{m,m\pm 2}$, that {\em vary slowly
with position}.\footnote{For the Hamiltonian~(\ref{ham0}), the hopping connects
sites differeing by $\Dta m = 2$, so that the eigenvalue problem divides into
two subspaces depending on whether $S-m$ is even or odd. This point is of no
consequence for application of the discrete WKB method itself, as the problem
can be recast as one of nearest-neighbor hopping in each subspace.} (See Fig.~2.) 
\begin{figure}
\centerline{\psfig{figure=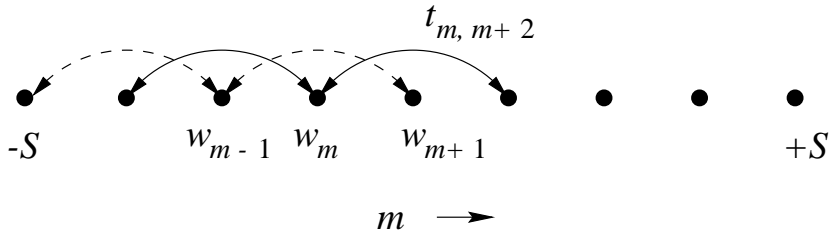}}
\caption{Mapping of spin problem onto an electron hopping on a lattice. The diagram
is drawn for a case such as Eq.~(\ref{ttr}) with only second neighbour hopping.
It is evident that the sites connected by dashed and solid lines
then belong to two disjoint subspaces of the Hamiltonian.}
\end{figure}
This viewpoint immediately suggests the approximation of semiclassical electron
dynamics, and indeed this approximation is identical to discrete WKB.

To apply semiclassical dynamics, we define local $m$-dependent
functions $w(m)$ and $t(m)$ by
\beq
\begin{array}{rcl}
w(m) & = & w_m, \\
t(m) & = & (t_{m,m+2} + t_{m,m-2})/2,
\end{array}
\label{wtcon}
\eeq
which we extend to continuous values of $m$ by demanding that they be smooth,
and that $dw/dm$ and $dt/dm$ be of relative order $S^{-1}$ with $m$ formally
regarded
as a quantity of order 1. A particle of energy $E$ can be assigned a local,
$m$-dependent, wavevector $q(m)$ in complete analogy with the continuous
WKB approach. The only difference is that the kinetic energy is given by
$2t(m) \cos q(m)$ instead of $q^2(m)/2\mu$ ($\mu$ being the mass). We thus
obtain
\beq
q(m) = \cos^{-1} \left({E - w(m) \over 2 t(m)} \right).
\label{qloc}
\eeq
The general solution to (\ref{ttr}) is given by linear combinations of
\beq
a_m \sim {1\over \sqrt{v(m)}} \exp \left(\pm i\int^m q(m') {dm' \over 2}\right).
\label{cmeq}
\eeq
In this equation,
\beq
v(m) = -2t(m) \sin q(m)
\label{vel}
\eeq
is the local particle velocity (equal to $\ptl E/\ptl q$), and the factor
$[v(m)]^{-1/2}$ fixes the normalization so that the probability current
is conserved. Also, the factor of $1/2$ in the integrand reflects the
step length $\Dta m = 2$ in Eq.~(\ref{ttr}).

We conclude here our general discussion of the discrete WKB method, and continue the
illustration with the Hamiltonian (\ref{ham0}) in the next section.
Braun's review~\cite{pab} contains a proper proof of the above aprroximations,
as well as connection formulas at turning points, Bohr-Sommerfeld quantization
rules, etc. Most of these are physically apparent, and the chief novelty arises
from the fact that for particle in a periodic potential (or a tight-binding
model), the allowed energies lie in a {\em band}, and are bounded both above
and below. This gives rise to turning points when the energy equals the local
{\em upper} band edge $w(m) + 2|t(m)|$, in addition to those that occur when
the energy equals $w(m) - 2|t(m)|$. These features are illustrated in Fig.~3.
\begin{figure}
\centerline{\psfig{figure=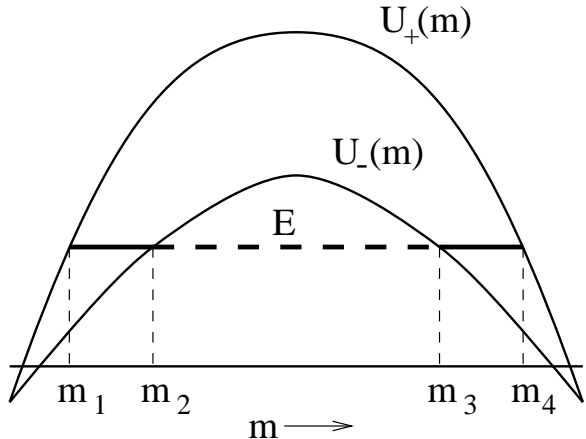}}
\caption{Discrete WKB method potential energy diagram for a symmetric double-well
problem. The functions $U_{\pm}(m) = w(m) \pm 2 t(m)$ are the local band edges.
For a particle with energy $E$ as shown, the regions $m_1 \le m \le m_2$,
and $m_3 \le m \le m_4$ are classically accessible, and represent the two wells.
The particle must tunnel across the classically forbidden region between
$m_3$ and $m_4$.}
\end{figure}
Braun refers to the first kind of turning
point as ``unusual", but the associated formulas are easily derived from those
for a ``usual" turning point by making a gauge transformation which changes the
sign of every other coefficient $a_m$. The associated bookkeeping can be fairly
cumbersome, however, and it is easy to make mistakes.

\section{Tunnel Splitting for the Hamiltonian~(\ref{ham0})}
We now turn to applying the discrete WKB method to finding the ground state
wavefunction and tunnel splitting $\Dta_0$ for the Hamiltonian (\ref{ham0}).
(We can also find the higher splttings and wavefunctions.~\cite{agunp}) 
There are three steps in finding the wavefunction itself: (i) find it in the
classically allowed region, (ii) find it in the
classically forbidden region $|m| \ll S$, (iii) match the two parts of the
wavefunction using the connection formulas or otherwise. The splitting is
then obtainable by a textbook formula.\footnote{See Eq.~(\ref{CH}) below.
This formula is anologous to one that appears in
the solution to Problem 3, Sec. 50, in the famous text by Landau and
Lifshitz.\cite{LL} However, it is {\em not} the final formula in that solution!
We have found it preferable to use the unnumbered intermediate
equation in which the splitting
is given as a product of the wavefunction and its derivative at the symmetry point
of the potential. The three dimensional analog of the latter formula is often named
after C. Herring who first used it find the splitting of the electron terms
in the H$_2^+$ ion. This problem is also discussed by Landau and Lifshitz
in the solution to the problem accompanying Sec. 81.}

To find the wavefunction in the classically allowed region near $m=-S$, e.g.,
it is advantageous to write $C_n = a_{-S +n}$, with $n=0, 1, \ldots$. It is
generally necessary to obtain the functions $w(m)$ and $t(m)$ to an accuracy
such that the first two terms in an expansion in powers of $S^{-1}$ are correctly
given; if only the leading term is kept, then retention of the
$\bigr(v(m)\bigr)^{-1/2}$ factor in Eq.~(\ref{cmeq}) can not be justified. In carrying
out this exercise near the ends of the chain, as is the case now, it is
necessary to treat $n = S+m$ rather than $m$ as a quantity of order $S^0$. It is
also useful to add a constant $k'_1 S(S+1)$ to the Hamiltonian. The Schr\"odinger
equation then reads
\beq
(2k_1 + k_2)\left(n + \half\right) C_n + \half k_2\sqrt{n(n-1)} C_{n-2}
      + \half k_2\sqrt{(n+1)(n+2)} C_{n+2} = E C_n.
\label{ttrho}
\eeq
This equation, however, can be solved exactly. It is that for the harmonic
oscillator Hamiltonian
\beq
\ham_{\rm ho} = (2k_1 + k_2) \adag a + \half k_2 (a^2 + {\adag}^2),
\label{hho}
\eeq
where $\adag$ and $a$ are raising and lowering operators for the number eigenstates
$\ket n$ obeying $\adag a \ket n = n \ket n$, and $C_n = \tran n\psi$ for an energy
eigenfunction $\ket\psi$. We can diagonalize $\ham_{\rm ho}$ by writing
$a = (x + i p)/2^{1/2}$, $\adag = (x - i p)/2^{1/2}$, where $x$ and $p$ are
canonical position and momentum operators obeying $[x,p] = i$ as usual.
This yields
\beq
\ham_{\rm ho} = k_{12} x^2 + k_1 p^2.
\label{hhox}
\eeq
The wavefunction $C_n$ (automatically normalized to unity) is now easily
obtained by evaluating the overlap $\tran n\psi$ in the $x$ representation.
We quote the final result for the ground state. $C_n$ vanishes for odd $n$,
and for even $n$ it equals
\beq
C_n = \sqrt{\sech\tta} {n \choose n/2}^{1/2}
           \left( -{\tanh\tta \over 2}\right)^{n/2}, 
\label{cpn}
\eeq
where
\beq
\tta = (1/4)\ln (k_{12}/k_1).
\label{tta}
\eeq
It is also useful to note the relations $\sinh 2\tta = k_2/\om_e$,
$\cosh 2\tta = (2k_1 + k_2)/\om_e$, and $\tanh 2\tta = k_2/(2k_1 + k_2)$,
with $\om_e = 2(k_1 k_{12})^{1/2}$ as defined in Eq.~(\ref{fel}). That
$\om_e$ is the small precession frequency for the electronic moments
is now obvious from Eq.~(\ref{hhox}).

[We mention for completeness that the above results also follow from
making a Bogoliubov transformation
\beq
\begin{array}{rl}
{}& a  =  \cosh\tta \ b - \sinh\tta \ \bdag,\\
{}& \adag  =  -\sinh\tta \ b + \cosh\tta\ \bdag, \\
\end{array}
\label{Bog}
\eeq
which, with $\tta$ given by Eq.~(\ref{tta}), diagonalizes $\ham_{\rm ho}$.
The coefficients $C_n$ follow from an obvious {\em two}-term recursion relation
connecting $C_n$ and $C_{n-2}$.]

The only constraint used in deriving Eq.~(\ref{cpn}) was $n\ll S$. In particular,
the result holds for $S \gg n \gg 1$. Thus it actually
extends out into part of the classically
forbidden region. This is a great advantage as it allows us to match to the
quasiclassical form of the wavefunction on the other side of the turning point
directly, obviating the need for connection formulas.

The wavefunction in the classically forbidden region is directly found
using the formulas (\ref{cmeq}) and (\ref{vel}). Under the conditions
$n/2 \gg 1$, $S -n/2 \gg 1$, we find that for even $n$
\beq
C_n = \left( \pi{n\over 2}\left(1- {n \over 2S}\right)^3 \right)^{-1/4}
              ({\sech\tta})^{1/2} (-\tanh\tta)^{n/2}.
\label{cnqc}
\eeq
($C_n$ continues to vanish for odd $n$.)
We have already adjusted the multiplicative
constant in this result, so as to agree with Eq.~(\ref{cpn}) when $S \gg n \gg 1$.

The tunnel splitting is generally given by the formula (see footnote {\it f\/})
\beq
\Dta_0 = \cases{
        2 t_{2,0} |a_0 (a_2 - a_{-2})|, & even $S$,\cr
        2 t_{1, -1}|a^2_1 - (a_{-1})^2|, & odd $S$. \cr}
\label{CH}
\eeq
(It is also easily shown that the antisymmetric state is lower in
energy for odd $S$, and higher for even $S$.)
Applying this formula to the problem at hand, we obtain 
\beq
\Dta_0 = 4\om_e \left( {S \over \pi} \right)^{1/2} \sech\tta \tanh^S \tta.
\label{Dtaex}
\eeq
Since $\tanh\tta < 1$, the last factor in this result can be rewritten
as a WKB or Gamow factor $\exp(-S |\ln\tanh\tta|)$. For $k_1/k_{12} \ll 1$,
this is approximately equal to $\exp(-4S k_{12}/\om_e)$, which is of the same
form as Eq.~(\ref{spapr}).

\begin{table}[t]
\caption{Comparison between numerical and analytical [Eq.~(\ref{Dtaex})]
results for the tunnel splitting without nuclear spins. The parameters
are $k_1 = 5.0$, $k_2 = 20.0$.\label{tab1}}
\vspace{0.2cm}
\begin{center}
\begin{tabular}{|c c c c|}
\hline
$S$ & $\Dta_0$ (numerical) & $\Dta_0$ (analytic) & Error(\%) \\
\hline
10 & $9.282 \times 10^{-3}$ & $9.749 \times 10^{-3}$ & 5.0 \\
11 & $3.738 \times 10^{-3}$ & $3.906 \times 10^{-3}$ & 4.5 \\
12 & $1.497 \times 10^{-3}$ & $1.558 \times 10^{-3}$ & 4.1 \\ 
13 & $5.974 \times 10^{-4}$ & $6.195 \times 10^{-4}$ & 3.7 \\
14 & $2.375 \times 10^{-4}$ & $2.455 \times 10^{-4}$ & 3.4 \\
15 & $9.412 \times 10^{-5}$ & $9.708 \times 10^{-5}$ & 3.1 \\
16 & $3.721 \times 10^{-5}$ & $3.830 \times 10^{-5}$ & 2.9 \\
17 & $1.468 \times 10^{-5}$ & $1.508 \times 10^{-5}$ & 2.7 \\
18 & $5.778 \times 10^{-6}$ & $5.927 \times 10^{-6}$ & 2.6 \\
19 & $2.271 \times 10^{-6}$ & $2.326 \times 10^{-6}$ & 2.4 \\
20 & $8.910 \times 10^{-7}$ & $9.115 \times 10^{-7}$ & 2.3 \\
\hline
\end{tabular}
\end{center}
\end{table}              

It is of interest to note that the only previous calculations of $\Dta_0$ to
the same accuracy (and with which we agree completely) are by Enz and
Schilling,~\cite{ES} and by Belinicher, Provdencia, and Providencia,~\cite{Bel}
both of whom employ methods of much greater complexity. Other tunnel splitting
calculations either find an incorrect prefactor,~\cite{vHS,gk}
or not at all.~\cite{cg} In Table 1, we compare the formula with exact answers
for the splitting obtained by numerical diagonalization of $\ham_0$
for $S$ ranging from 10 to 20,
for parameters $k_2 = 20.0$, and $k_1 = k_2/4 = 5.0$. This choice yields
$\om_e = 22.36$, $\tanh\tta = 0.3820$, and $\sech\tta = 0.9242$. As can be
seen, the errors are quite small and the results improve steadily with
increasing $S$. Note further that Eq.~(\ref{Dtaex}) consistently overestimates
the splitting. These results are in keeping with previous numerical
studies.~\cite{ES,Bel}

\section{Inclusion of Nuclear Spins}
\subsection{One-Coflip Splittings}
The results of the previous section enable us to calculate the tunnel splitting
$\Dta(I,p/2)$ for small values of $I$ and $p$. Let us begin with the simplest
possible case, when $I=1/2$, and $p=\pm 1$. Then as discussed in Sec. 2, it is
necessary to take $S$ to be a half integer. We will write $S = N + 1/2$, where
$N$ is an integer.

In the language developed in the last section, the problem is now
equivalent to that of an electron hopping on a lattice with {\em two} rows,
corresponding to $I_z = \pm 1/2$, and $2S + 1 = 2(N+1)$ lattice points per row.
The problem is represented pictorially in Fig.~4, where the various non-zero
off-diagonal elements of the Hamiltonian (\ref{hamI}) are shown as bonds connecting
the corresponding lattice points.\
\begin{figure}
\centerline{\psfig{figure=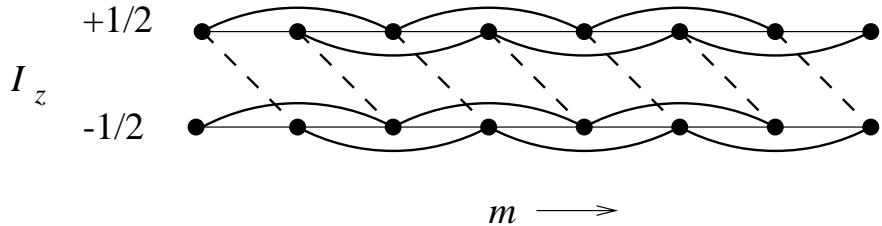}}
\caption{Lattice diagram for MQC problem with one nuclear spin coflip. The
$k_xS_x^2$ and $S_{\pm}I_{\mp}$ bonds are shown by heavy solid and dashed lines
respectively. The division into two subspaces is again eident.}
\end{figure}
The bonds connecting alternate sites in the
same row arise from the $k_2 S_x^2$ term, while the bonds connecting sites in
different rows arise from the terms $S_+I_-$ and $S_-I_+$. It is apparent that the
sites form two groups that are disconnected from each other, or in proper
mathematical language, that the Hamiltonian divides into two disjoint subspaces.
In particular the splitting of the two lowest states in the space that contains
the state\footnote{It is necessary in this section to refer to
two different types of states with similar labels: the states $\ket{S_z,I_z}$,
and the states $\ket{\pm n, p}$, where $n$ is the harmonic-oscillator-like
excitation index used in Sec.~2.2. We shall avoid notational confusion by refering
to the second type of state as $\ket{0,p=1}$, $\ket{-0,p=1}$, etc., in which
the label $p$ is explicitly identified.} $\ket{-S,-1/2}$ 
equals $\Dta(1/2,1)$, while the splitting in the
space containing $\ket{-S,+1/2}$ equals $\Dta(1/2,-1)$.

Let us first calculate $\Dta(1/2,-1)$ [which equals $\Dta_{-1}$ as we will
show in Eq.~(\ref{Dip}) below]. The lattice
diagram for this is shown in Fig.~5, where
we have thrown away the states in the other subspace, and we have relabeled
the sites as explained below.
\begin{figure}
\centerline{\psfig{figure=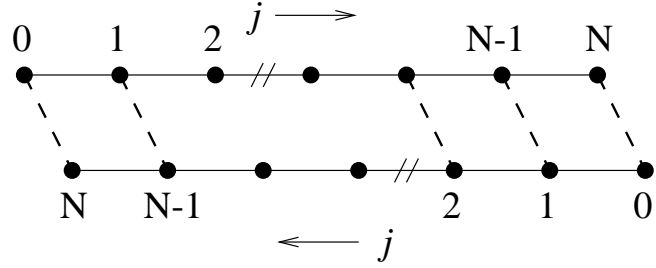}}
\caption{Lattice diagram and labelling scheme for calculation of overlap for
$\Dta(1/2,-1)$. $N= S-1/2$. Only sites in the subspace including $S_z = -S$,
$I_z = 1/2$ are shown.}
\end{figure}
We will obtain the
splitting by finding a suitable effective two-state Hamiltonian. To this end, let
us first ignore the $S_+I_-$ and $S_-I_+$ terms in $\ham_I$. The two lowest
energy states are then $\ket{+0,p=-1} \equiv \kI$ and
$\ket{-0,p=+1} \equiv \kII$. These states live on
different rows of the lattice, with $\kI$ and $\kII$ being
localized near the right edge of the lower row, and the 
left edge of the upper row, respectively. It should be carefully noted that neither
state is large near both ends. The state which is large near $S_z= -S +1$ and
$I_z = -1/2$ (lower left corner), for example, is approximately equal to
$\ket{-1,p=-1}$ which is higher in energy than $\kI$ or $\kII$
by $\om_e$. The additional bias in
the on-site energy due to the term $\om_n S_z I_z/S$ is not large enough to offset
this energy difference since $\om_n \ll \om_e$. For the same reason, the wavefunctions
of the states $\kI$ and $\kII$ are very well approximated by the wavefunction
found in the previous section. It is necessary only to redefine the position coordinate
appropriately. We do this by labelling the sites in Fig.~5 by $j = 0, 1, \ldots, N$,
where $j$ runs from left to right for the upper row, and in
the opposite direction for the lower row. Then
\beq
\tran{S-2j,-1/2}{{\rm I}} = \tran{-S+2j,+1/2}{{\rm II}} \equiv C'_j.
\label{cprj}
\eeq
Further, $C'_j = C_{2j}$ where $C_{2j}$ is given by Eqs.~(\ref{cpn}) and (\ref{cnqc}).

The effective two-state Hamiltonian can be written as
\beq
\ham_{\rm eff} = \left( \matrix{E_0 & V \cr
                                V & E_0 \cr} \right),
\label{heff}
\eeq
where $E_0 = \bI{\ham_I}\kI = \bII{\ham_I}\kII$, and
$V = \bI{\ham_I}\kII = \bII{\ham_I}\kI$. The value of $E_0$ is clearly
immaterial, while the tunnel splitting is given by $2|V|$. The off-diagonal
element $V$ arises from the $S_+I_-$ and $S_-I_+$ terms and can be directly
evaluated because $\om_n \ll \om_e$.  We have
\bea
V & = & \sum_{j=0}^N C'_j C'_{N-j}
   \left\langle -S+2j+1,-\half \biggl| -{\om_n\over 2S} S_+I_-
         \biggr| -S+2j, \half \right\rangle,  \nnu\\
                 & = & -(\om_n/2S)\sum_{j=0}^N C'_j C'_{N-j}
                          \left[(2j + 1)(2S - 2j)\right]^{1/2}.
\label{ofd}
\eea
The dominant contribution to the sum arises from the middle of the chains, i.e.,
from values of $j$ such that $(N-j), j \gg 1$. Using Eq.~(\ref{cnqc}) and
neglecting terms of order unity in comparison with $j$ and $N-j$, we obtain
\beq
V = -(-1)^N 2\,\om_n (S/\pi)^{1/2} \sech\tta \tanh^N\tta
                       \sum_j\left[(j +1/2)(N-j+1/2)\right]^{-1/2}.
\label{ofd2}
\eeq
For $N \gg 1$ the sum may be approximated by an integral, which is easily shown
to equal $\pi$. Making use of Eq.~(\ref{Dtaex}), and recalling
that $S = N + 1/2$, we can write the tunnel splitting (which equals $2|V|$) as
\beq
\Dta_{-1} = \Dta(1/2,-1) = {\pi\over 2}{\om_n\over\om_e}(\tanh\tta)^{-1/2}\Dta_0.
\label{Dtaaf}
\eeq

The splitting $\Dta(1/2,1)$ ($= \Dta_1$) can be found in the same way. The appropriate
lattice diagram is drawn in Fig.~6.
\begin{figure}
\centerline{\psfig{figure=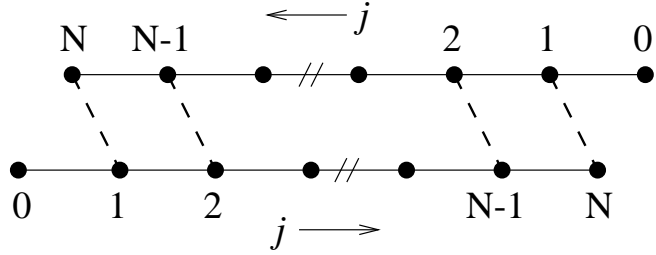}}
\caption{Same as Fig.~5 for $\Dta(1/2, 1)$. Only sites in the subspace $S_z=S$,
$I_z = 1/2$ are shown this time.}
\end{figure}
This time the two low energy states are
$\ket{+0,p=1}$ and $\ket{-0,p=-1}$, and the wavefunctions are given by the same
$C'_j$ as above with the sites labeled as shown. The key quantity is again the
off-diagonal element. It is obvious from the figure that this is again given by
a sum as in Eq.~(\ref{ofd}). The only change required is that we replace
the product of wave functions in that summand by $C'_j C'_{N+1 -j}$. The sum
is again dominated by the central region, and since
$C'_{j+1} \approx -\tanh\tta C'_j$ in this region, the change in the summand
has the effect of multiplying the each term in the sum by $-\tanh\tta$. The
final result for the splitting is, therefore,
\beq
\Dta_1 = \Dta(1/2,1) = {\pi\over 2}{\om_n\over\om_e}(\tanh\tta)^{1/2}\Dta_0.
\label{Dtaf}
\eeq

We thus see that apart from the factors of $(\tanh\tta)^{\pm 1/2}$, the results
(\ref{Dtaaf}) and (\ref{Dtaf}) agree with the instanton answer, Eq.~(\ref{Delpin}).
This agreement shows that the instanton method is basically sound, and although
it can not in its simplest version give the prefactors in $\Dta_0$ correctly, the
overall picture it provides is correct.
In fact, the $(\tanh\tta)^{\pm 1/2}$ factors can also be found
from the instanton approach by exercising
a little more care. These origin of these factors is actually very easy to understand.
They reflect the fact that the state with antiparallel alignment of the electronic
and nuclear spins has greater quantum fluctuations or zero point motion than the
state with parallel alignment. Another way to say this is that the state
$\ket{S_z=S,I_z=I}$ is an eigenstate of the hyperfine coupling term in $\ham_I$,
while $\ket{S_z=S,I_z=-I}$ is not. This difference in zero point fluctuations
shows up as a higher tunneling rate for coflips in which the nuclear spins
are oppositely aligned to the electronic ones.

\begin{table}[t]
\caption{Comparison between numerical and analytical [Eqs.~(\ref{Dtaaf})
and (\ref{Dtaf})] results for one-coflip tunnel splitting. The parameters
are $k_1 = 5.0$, $k_2 = 20.0$, and $\om_n = 0.1$. For each value of $S$,
the entry in the upper row is $\Dta_1$, and that in the lower row $\Dta_{-1}$.
\label{tab2}}
\vspace{0.2cm}
\begin{center}
\begin{tabular}{|c c c c|}
\hline
$S$ & $\Dta_{\pm 1}$ (numerical) & $\Dta_{\pm 1}$ (analytic) & Error(\%) \\
\hline
$10\half$ & $\efrm{2.576}{5}$ & $\efrm{2.681}{5}$ & 4.1 \\
          & $\efrm{6.661}{5}$ & $\efrm{7.018}{5}$ & 5.4 \\
\hline
$11\half$ & $\efrm{1.034}{5}$ & $\efrm{1.072}{5}$ & 3.6 \\
          & $\efrm{2.674}{5}$ & $\efrm{2.805}{5}$ & 4.9 \\
\hline
$12\half$ & $\efrm{4.134}{6}$ & $\efrm{4.267}{6}$ & 3.2 \\
          & $\efrm{1.069}{5}$ & $\efrm{1.117}{5}$ & 4.5 \\
\hline
$13\half$ & $\efrm{1.646}{6}$ & $\efrm{1.694}{6}$ & 2.9 \\
          & $\efrm{4.257}{6}$ & $\efrm{4.435}{6}$ & 4.2 \\
\hline
$14\half$ & $\efrm{6.534}{7}$ & $\efrm{6.705}{7}$ & 2.6 \\
          & $\efrm{1.689}{6}$ & $\efrm{1.755}{6}$ & 3.9 \\
\hline
$15\half$ & $\efrm{2.587}{7}$ & $\efrm{2.648}{7}$ & 2.4 \\
          & $\efrm{6.688}{7}$ & $\efrm{6.933}{7}$ & 3.7 \\
\hline
$16\half$ & $\efrm{1.022}{7}$ & $\efrm{1.044}{7}$ & 2.2 \\
          & $\efrm{2.641}{7}$ & $\efrm{2.732}{7}$ & 3.5 \\
\hline
$17\half$ & $\efrm{4.024}{8}$ & $\efrm{4.105}{8}$ & 2.0 \\
          & $\efrm{1.041}{7}$ & $\efrm{1.075}{7}$ & 3.3 \\
\hline
$18\half$ & $\efrm{1.583}{8}$ & $\efrm{1.612}{8}$ & 1.9 \\
          & $\efrm{4.092}{8}$ & $\efrm{4.221}{8}$ & 3.2 \\
\hline
$19\half$ & $\efrm{6.217}{9}$ & $\efrm{6.322}{9}$ & 1.7 \\
          & $\efrm{1.608}{8}$ & $\efrm{1.655}{8}$ & 3.0 \\
\hline
$20\half$ & $\efrm{2.438}{9}$ & $\efrm{2.476}{9}$ & 1.6 \\
          & $\efrm{6.303}{9}$ & $\efrm{6.482}{9}$ & 2.9 \\
\hline
\end{tabular}
\end{center}
\end{table}              

We also remind readers that in Eqs.~(\ref{Dtaaf}) and (\ref{Dtaf}), $\Dta_0$ is not
the tunnel splitting for the bare problem without nuclear spins (which vanishes
because $S$ is half-integral), but rather the pseudo-splitting formally given
by Eq.~(\ref{Dtaex}).

In Table 2, we show the comparison between the above formulas and exact
answers from numerical diagonalization of $\ham_I$ with $I=1/2$.
We choose $k_1 =5.0$ and $k_2 = 20.0$
as before, and $\om_n = 0.1$. Since $\om_e = 22.36$, the condition
$\om_n \ll \om_e$ is well obeyed. Again the agreement is very satisfactory, and
improves with increasing $S$. In particular, the inclusion of the
$(\tanh\tta)^{\pm 1/2}$ factors seems to be required.

\subsection{Two-Coflip Splittings}
Let us proceed to calculate the two-coflip splittings, $\Dta(1,\pm 2)$
($=\Dta_{\pm 2}$), by studying the Hamiltonian (\ref{hamI}) for $I=1$ and
integer $S$. As before, we carry out the calculation by including the $S_zI_z$ term
in the unperturbed Hamiltonian along with $\ham_0$, and treating the $S_+I_-$
and $S_-I_+$ terms as a perturbation. 

Suppose we wish to find $\Dta(1,2)$. The appropriate
lattice diagram is shown in Fig.~7.
\begin{figure}
\centerline{\psfig{figure=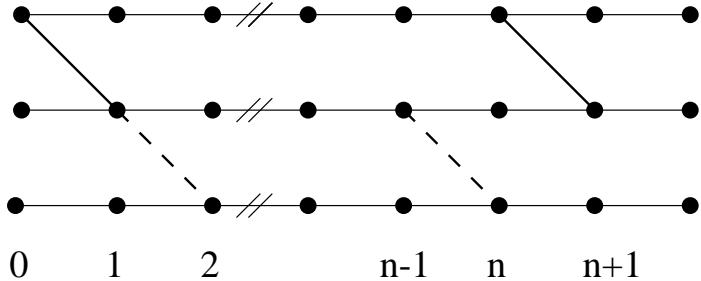}}
\caption{Lattice diagram for the case $I=1$. The bonds entering into the
calculation of the transition matrix element $V$ relevant to $\Dta(1,2)$
and $\Dta(1,-2)$ are shown as dashed and solid diagonal lines respectively.}
\end{figure}
This tunneling process requires making a transition from the state
$\ket{-0,p=-2}\equiv\kI$ to the state $\ket{0,p=2}\equiv\kIV$. However,
there is no direct matrix element of $\ham_I$ between these states, and we
must go through intermediate states. The lowest energy intermediate states are
$\ket{-1,p=0}\equiv\kII$ and $\ket{1,p=0}\equiv\kIII$.
Keeping only these we obtain a four-state effective Hamiltonian
\beq
\ham_{\rm eff} = \left( \matrix{E_0 & V & 0 & 0 \cr
                                V   & E_1 & \Dta^{(1)}/2 & 0 \cr
                                0   & \Dta^{(1)}/2 & E_1 & V \cr
                                0   & 0 & V & E_0 \cr} \right),
\label{h4b4}
\eeq
where the equality of various elements is assured by symmetry. Here,
$E_0 = \bI{\ham_I}\kI = \bIV{\ham_I}\kIV$,
$E_1 = \bII{\ham_I}\kII = \bIII{\ham_I}\kIII$,
$V = \bII{\ham_I}\kI$, and $\Dta^{(1)}$ is the tunnel splitting between the
states $\kII$ and $\kIII$. This splitting is identical to the splitting between
the first excited states of the Hamiltonian $\ham_0$ without nuclear spins, and is
given by~\cite{agunp}
\beq
\Dta^{(1)} = {2S\over \sinh\tta \cosh\tta}\Dta_0.
\label{Dta1}
\eeq
Secondly, the energy difference $E_1 - E_0 = \om_e +{\cal O}(\om_n)$.

The key quantity in calculating $\Dta_2$ is thus the matrix element $V$
as before. This is given by a sum involving the lattice site wavefunctions
of the states $\kI$ and $\kII$ and a $I_-S_+$ matrix element. Refering
to Fig.~7, consider the contribution of the bond connecting site $n$ in the 
lower row to site $n-1$ in the middle row. The matrix element for this bond
equals
\beq
- {\om_n\over 2S} \mel{-S+n -1,I_z =0}{S_+I_-}{-S+n,I_z = -1}
  = - {\om_n\over \sqrt2 S}\bigl( n(2S - n + 1) \bigr)^{1/2}.
\label{bond}
\eeq
The wavefunctions at the two ends of this bond, on the other hand, are given by
$\tran n{\psi_0}$ and $\tran {n-1}{\psi_1}$, where $\ket{\psi_0}$
and $\ket{\psi_1}$ are the
ground and first excited state of the harmonic oscillator Hamiltonian $\ham_{\rm ho}$,
Eq.~(\ref{hho}), and $\ket n$ are the eigenstates of the number operator $\adag a$.
Thus,
\bea
V  &=& -{\om_n\over \sqrt2 S}\sum_{n =0}^{\infty}  
        \bigl( n(2S - n + 1) \bigr)^{1/2}\tran{\psi_1}{n-1} \tran{n}{\psi_0} \nnu \\
   &=& -{\om_n \over \sqrt S} \sum_{n=0}^{\infty}
        n^{1/2}\tran{\psi_1}{n-1} \tran{n}{\psi_0} \nnu \\
   &=& -{\om_n \over \sqrt S} \sum_{n=0}^{\infty}
            \mel{\psi_1}an \tran n{\psi_0} \nnu \\
   &=& -{\om_n \over \sqrt S} \mel{\psi_1}a{\psi_0} \nnu \\
   &=& {\om_n \over \sqrt S}\sinh\tta.
\label{Vdta1}
\eea
The second equality in this chain follows from noting that the sum is dominated
by values of $n$ of order unity, which allows us to neglect $n$ in comparison to
$S$. The final equality follows if one makes use of the Bogoliubov transformation
(\ref{Bog}). Then, $\mel{\psi_1}a{\psi_0} =
           -\sinh\tta \mel{\psi_1}\bdag\psi_0 = -\sinh\tta$.

To leading order in $\om_n$, the energy splitting of the lowest two states
of the effective Hamiltonian (\ref{h4b4}) is easily shown to be given by
\beq
 {V^2 \over E_1 - E_0 - \Dta^{(1)}/2} - {V^2 \over E_1 - E_0 + \Dta^{(1)}/2} 
 \approx {V^2 \over (E_1 - E_0)^2} \Dta^{(1)}.
\label{dta2temp}
\eeq
Combining Eqs.~(\ref{Dta1}), (\ref{Vdta1}), and using the fact that 
$E_1 - E_0 = \om_e +{\cal O}(\om_n)$, we finally obtain
\beq
\Dta_2 = \Dta(1,2) \approx 2{\om_n^2 \over \om_e^2} \tanh\tta \Dta_0.
\label{Dta2f}
\eeq

The calculation of $\Dta(1,-2)$ proceeds very similarly. The effective Hamiltonian
has the same structure as Eq.~(\ref{h4b4}), except that this time the states
$\kI$ and $\kIV$ must be taken as $\ket{-0,p=2}$ and $\ket{+0,p=-2}$
respectively, while the states $\kIII$ and $\kIV$ are unchanged. This time
the bonds connect site $n$ in the upper row with site $n+1$ in the lower row, and
the matrix element is
\beq
- {\om_n\over 2S} \mel{-S+n +1, I_z =0}{S_-I_+}{-S+n,I_z = 1}
  \approx - {\om_n\over \sqrt S} (n + 1)^{1/2}
\label{bond2}
\eeq
instead of Eq.~(\ref{bond}). Using the same technique as before, the transition
matrix element $V$ is found to be 
\beq
V =  -{\om_n \over \sqrt S} \sum_{n=0}^{\infty}
        (n+1)^{1/2}\tran{\psi_1}{n+1} \tran{n}{\psi_0} 
   = -S^{-1/2} \om_n \cosh\tta.
\label{Vdta2}
\eeq
Substituting this in Eq.~(\ref{dta2temp}), we obtain the splitting as
\beq
\Dta_{-2} = \Dta(1,2) \approx 2{\om_n^2 \over \om_e^2} (\tanh\tta)^{-1} \Dta_0.
\label{Dta2af}
\eeq

\begin{table}[t]
\caption{Comparison between numerical and analytical results for two-coflip
tunnel splittings, for $k_1 = 5.0$, $k_2 = 20.0$, $\om_n = 0.5$, $S=15$,
and $I=1$.  In the column labeled `ratio' we give the ratio of the numerically
obtained splittings to those given by the instanton method, i.e., to
Eqs.~(\ref{Dta2f}) and (\ref{Dta2af}), but with a numerical factor of $\pi^2/4$
instead of 2.
\label{tab3}}
\vspace{0.2cm}
\begin{center}
\begin{tabular}{|l c c|}
\hline
& Numerical & Ratio  \\
\hline
$\Dta_{2}$  & $\efrm{2.559}{8}$  & 0.56  \\
$\Dta_{-2}$ & $\efrm{3.488}{7}$  & 1.11 \\
\hline
\end{tabular}
\end{center}
\end{table}

The results (\ref{Dta2f}) and (\ref{Dta2af}) are very close to but not exactly 
what one would expect from the instanton method, even after the fluctuational
factors of $\tanh\tta$ have been included. If we define
$\eta_{\pm} = (\pi\om_n/2\om_e)(\tanh\tta)^{\pm 1/2}$, then an instanton argument
would lead us to expect
\beq
\Dta_{\pm 2} = \eta^2_{\pm} \Dta_0.
\label{insex}
\eeq
This would require the numerical factors in Eqs.~(\ref{Dta2f})
and (\ref{Dta2af}) to be $\pi^2/4$ instead of 2. Could this really be the case,
and if so, what is the source of the discrepancy in our present calculation?
One possible answer is that we have ignored the contribution of higher energy
intermediate
states in the calculation of $\Dta_{\pm 2}$. In fact, it can be shown that these states'
contribution is also formally of order $(\om_n/\om_e)^2 \Dta_0$, but with different
numerical prefactors, which decrease rapidly with increasing intermediate state energy.
The accurate calculation of these numerical factors is difficult. Similarly, we neglected
the terms of higher order in $S^{-1}$ in expanding the matrix elelment (\ref{bond}).
Terms of such higher order are also present in the contributions from the higher
intermediate states, and it is not clear that the {\em sum} of all these terms will
continue to be of higher order in $S^{-1}$. When these approximations
are taken into account, we cannot exclude the possibility that Eq.~(\ref{insex})
holds exactly.

We show a comparison between our theoretical and numerical results in Table 3.
The quality of
agreement is still fairly good, though not as impressive as for the one-coflip 
splittings. We have not carried out this numerical work as extensively as in the
previous cases, and so can not comment on the behavior with increasing $S$.

\subsection{Higher Coflip Processes; Effective Hamiltonian}
It is evident that the discrete WKB method is increasingly ill suited to
the calculation of higher coflip processes. Systematic numerical
investigation of this problem runs into the difficulty that the splittings
necessarily decrease with increasing coflip number $p$, requiring the
numerical diagonalization to be carried out to increasing precision.
My own numerical studies are very limited, and while it is undoubtedly
possible to improve on them, that would require substantially greater time and
effort than I am prepared to commit!

There is, however, one aspect of the higher coflip problem, which can be
understood quite simply. [We have alluded to this earlier; see the discussion
immediately preceding Eq.~(\ref{splbi}).] This is that splittings $\Dta(I,p)$
for different $I$ but equal $p$ can be related to each other. To see this, let
us consider the problem in terms of $N_n$ nuclear spins each with spin $1/2$
rather than the model (\ref{hamI}). There are then a large number of states
in the $\ket{+0,p}$ group that are degenerate with one another, and with an
equally large number of mutually degenerate states in the $\ket{-0,-p}$ group.
(Recall that $p = 2I^{\rm tot}_z$.) A $p$-coflip process can take us from any
of the states in the first group to several states in the second group.
A convenient algebraic method for keeping track of which states are connected
is as follows. To avoid cluttering up the formulas, it is best to treat the cases
$p >0$ and $p < 0$ separately. Let us do the $p > 0 $ case first.
Let $\pi_{\pm p}$ be projection operators
onto the set of states with $I^{\rm tot}_z = \pm p/2$, irrespective of the value of
$I$. Further, let
\beq
\begin{array}{rcl}
\sig_+ = \ket{+0}\bra{-0}, \\
\sig_- = \ket{-0}\bra{+0}, \\
\label{sigpm}
\end{array}
\eeq
be transition  operators for the large electronic spin between the two states
involved in the MQC process. Finally, let $\Dta_p$ be the amplitude for the
transition between any two of the states in the groups with opposite $I_z$.
Then, an effective Hamiltonian that describes
all the coflip processes of order $p$ is
\bea
&\ham^{\rm cf}_p = \Dta_p\Big(\sig_+ \pi_p Q^+_p \pi_{-p} +{\rm h.c.} \Big);
\label{hcf} \\
&Q^+_p = \displaystyle{\mathop{{\sum\sum\cdots\sum}}_{j_1,\ j_2,\ldots, j_p}}
           \ I^+_{j_1} I^+_{j_2}\cdots I^+_{j_p},
\label{Qpl}
\eea
where $I^{\pm} = I_x \pm i I_y$ as usual, and the sum in Eq.~(\ref{Qpl}) is over
all distinct $p$-tuplets of indices chosen from the indices $1, 2, \ldots, N_n$.

Let us check that Eq.~(\ref{hcf}) does what it is supposed to. Consider the first
term on the right, and let it operate on a ket.
The projection operator $\pi_{-p}$ ensures that this term
is only relevant when we operate on a state with $I_z=-p/2$. The operator $Q^+_p$
then raises (or flips) $p$ nuclear spins leading to a state with $I_z = p/2$.
The projection operator $\pi_p$ in this term is added to ensure that the result is
sensible when we operate on a bra. The second term in (\ref{hcf}) ensures
hermiticity and describes processes in which $I_z$ is lowered.

It is now very easy, however, to find the $I$ dependence of the matrix
elements of $\ham^{\rm cf}_p$. We first note that because $(I^+_j)^2 = 0$ for any $j$,
we can write
\beq
Q^+_p = {1\over p!}\Big(I^+_1 + I^+_2 +\cdots + I^+_{N_n}\Big)^p
        = {1\over p!}(I^+)^p.
\label{Qpl2}
\eeq
The only non-zero matrix element of $\ham^{\rm cf}_p$ between states
of given $I$ is thus equal to
\beq
{\Dta_p \over p!}\left\langle I, m=\half p
                        \Bigl| (I^+)^p \Bigr|  I, m=-\half p \right\rangle =
                  \Dta_p {I +\half p \choose p}.
\label{hcf2}
\eeq

The case $p < 0$ is now easily treated by minor changes of notation.
We simply change $Q^+_p$ in the first
term in Eq.~(\ref{hcf}) to its hermitean adjoint $Q^-_p$, with the sum in
the analog of Eq.~(\ref{Qpl}) being
over all $|p|$-tuplets. Thus $Q^-_p = (I^-)^{|p|}/|p|!$ for $p<0$, and the
matrix element analogous to (\ref{hcf2}) is identically evaluated. The upshot
is that we can write
\beq
\Dta(I,p) = {I + \half|p| \choose |p|} \Dta_p
\label{Dip}
\eeq
for all $p$, positive or negative.

\begin{table}[t]
\caption{Numerical results for higher coflip tunnel splittings. 
The parameters are $k_1 = 5.0$, $k_2 = 20.0$, $\om_n=0.5$, $S=15$,
and $I = 0, 1, 2, 3$. The table shows $\Dta(I,p)$ divided by the combinatoric
factor in Eq.~(\ref{Dip}). This combinatoric number is shown next to the
result in parentheses.
\label{tab4}}
\vspace{0.2cm}
\begin{center}
\begin{tabular}{|r c c c c|}
\hline
$p$ & $I=0$ & $I=1$ & $I=2$ & $I=3$ \\
\hline
0 & $\efrt{9.412}{5}1$ & $\efrt{9.435}{5}1$ & $\efrt{9.479}{5}1$
                                                & $\efrt{9.546}{5}1$ \\
-2 & & $\efrt{3.488}{7}1$ & $\efrt{3.497}{7}3$ & $\efrt{3.507}{7}6$ \\
 2 & & $\efrt{2.559}{8}1$ & $\efrt{2.586}{8}1$ & $\efrt{2.628}{8}6$ \\
-4 & & & $\efrt{2.134}{9}1$ & $\efrt{2.138}{9}5$ \\
 4 & & & $\efrt{2.655}{11}1$ & $\efrt{2.580}{11}5$ \\
-6 & & & & $\efrt{1.735}{11}1$ \\
 6 & & & & $\efrt{3.624}{13}1$ \\
\hline
\end{tabular}
\end{center}
\end{table}              

We show in Tables 4 and 5 some numerical results for
tunnel splittings involving up to 6 nuclear spin coflips. Instead of tabulating
$\Dta(I,p)$ itself, we have divided it by the combinatorial factor
in Eq.~(\ref{Dip}). We expect that the resultant quantities will be independent
of $I$, and we can see from the tables that indeed they are. In all the cases, the
spread in the values of $\Dta_p$ so deduced is less than 2\%.

\begin{table}[h]
\caption{Same as Table 4 for odd numbers of coflips. Now $S=31/2$, and
$I = 1/2, 3/2, 5/2$. The other parameters are unchanged.
\label{tab5}}
\vspace{0.2cm}
\begin{center}
\begin{tabular}{|r c c c|}
\hline
$p$ & $I=1/2$ & $I=3/2$ & $I=5/2$  \\
\hline
-1 & $\efrt{3.265}{6}1$ & $\efrt{3.272}{6}2$ & $\efrt{3.284}{6}3$ \\
 1 & $\efrt{1.327}{6}1$ & $\efrt{1.326}{6}2$ & $\efrt{1.323}{6}3$ \\
-3 & & $\efrt{1.619}{8}1$ & $\efrt{1.623}{8}4$ \\
 3 & & $\efrt{2.305}{9}1$ & $\efrt{2.306}{9}4$ \\
-5 & & & $\efrt{1.136}{10}1$ \\
 5 & & & $\efrt{6.457}{12}1$ \\
\hline
\end{tabular}
\end{center}
\end{table}              

The fact that $\Dta(I,p)$ grows with $I$ may make one wonder if this might not be
a way to boost the resonance frequency for some of the tunneling processes. In fact, the
result (\ref{Dip}) only applies to the ideal case where the hyperfine couplings are
identical for all the nuclei. This limits the relevance of this result for actual systems
to small values of $p$ or $I$. Real particles typically possess some small spread in
these couplings. This spread spoils the degeneracy of all the states in the $\ket{0,p}$
group (and likewise for the states $\ket{-0,-p}$). When the spread starts to equal
$\Dta_p$, then the constructive interferences which give rise to the large
combinatoric factors in Eq.~(\ref{Dip}) are no longer possible. The spectral
weight in $\chi''$ is then shifted from the higher frequency peaks in a given
coflip group to lower frequencies. Secondly, we have neglected incoherent processes
involving the nuclear spins. MQC is destroyed by a single such process, and since the
likelihood that at leat one such process will occur in a given time grows linearly with
$N_n$, the number of nuclear spins, it is evident that it
does not pay to increase this number.

\section{Conclusions}
The main purpose of this article has been to verify previous calculations~\cite{ag1,ag3}
of the coflip tunneling frequencies without using instanton methods. We find
that by and large, the instanton calculations lead to correct answers.

Our discussion up to this point has treated the system as closed, with no means
for energy to flow in and out. If this were strictly true, this would imply that
$\chi''$ consisted of a sum of delta functions at the frequencies $\Dta(I,p)$. This
is of course an idealization. To complete the discussion of the problem, we must also
consider relaxation. If we put in broadening by hand, the 
qualitative picture of the $\chi''$ spectrum is as shown in Fig.~8.
\begin{figure}
\centerline{\psfig{figure=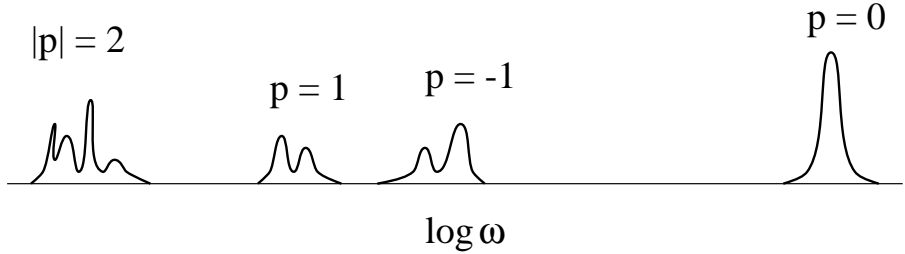}}
\caption{Sketch of the expected tunneling susceptibility spectrum $\chi''(\om)$
in the presence of nuclear spins, showing the first few polarization blocks. Note that
the frequency scale is logarithmic. We have tried to indicate the fine structure within
the higher polarization blocks, but we have made no attempt to represent the broadening
of the lines correctly.}
\end{figure}
A proper investigation of the physical mechanisms behind
this would take us into very different territory. Two obvious mechanisms which may
be mentioned in passing are phonons and dipolar magnetic fields due to other nuclei
such as protons which may be present in the material. It does not seem worthwhile to
develop a quantitative theory of the relaxation due to even these processes in the
absence of experimental impetus, but certain broadly valid comments can nevertheless
be made. By appealing to the general principles laid down in Refs.~11 and 29, it is
quite plausible that the relaxation can be treated by coupling the resonating variable
(which we would regard as different for each coflip line) to a phenomenological
harmonic oscillator bath (which would also be different for each coflip line). Broadly
speaking, such a bath has two effects. First, it leads to a pulling down or
downward renormalization of the bare coherence frequency $\Dta_b$ to ${\tilde\Dta}_b$.
At the simplest level this renormalization can be understood in terms of
a multiplicative Franck-Condon factor. Second, it gives rise to true damping of the
resonance, i.e., an intrinsic broadening or linewidth. (There may of course be
additional inhomogeneous broadening arising from a spread in ${\tilde\Dta}_b$.)
The general point which is noteworthy is
that as the bare frequency $\Dta_b$ of the coherence phenomenon under study decreases,
the environment suppresses it even further. This is
especially so if the environment has any subohmic or ohmic component. Environmental
degrees of freedom that might not have been relevant if $\Dta_b$ had been higher by a
factor of 10, say, gang up on the system, as it were, and do start to matter. Further,
degeneracy is broken by weaker and weaker stray fields and drifts, effectively
eliminating the resonance altogether.

All this implies that the higher coflip resonances in our problem are highly likely to
be overdamped and unobservable. This point can be seen even in the simplistic model
introduced in Ref.~9, and briefly touched upon
at the end of the previous section. If a single nuclear spin undergoes
an incoherent transition from $I_z = 1/2$ to $I_z = -1/2$ or vice versa at
a rate $1/\tau$, then any MQC in the particle as a whole is damped at a
rate $N_n/\tau$. All the lines in $\chi''(\om)$ thus acquire a width of order
$N_n/\tau$ irrespective of the number of coflips involved, and the higher coflip
lines with $\Dta(I,p) < N_n/\tau$ essentially merge into a mushy continuous
background with no clear feature that could be identified as a resonance. This
in turn means that while an accurate calculation of $\Dta(I,p)$ for large $|p|$
would still have theoretical interest, the unlikelihood of its relevance to
experiment reduces the urgency of this calculation considerably. The most interesting
remaining problem at this stage appears to be that of broadening of the very low
order coflip lines.

Let us therefore summarize our conclusions about
how MQC in small magnetic particles is affected by
nuclear spins. The effect is severe, and the spectrum of $\chi''$ is broken up into
a large number of lines which can be grouped by the number of nuclear spin coflips
required to maintain degeneracy, or more conveniently by the nuclear spin polarization
\beq
p = {2\over s}\sum_i s_{iz} I_{iz},
\label{pol}
\eeq
where $s_i$ and $I_i$ are the electronic and nuclear spins on atom $i$.
The frequency of a line decreases
quasi-geometrically with the number of coflips $|p|$, and there is a further
distinction between $p>0$ and $p<0$, with the latter having a higher frequency.
Within each polarization group, there is a fine structure to the tunneling
spectrum controlled by interference between the different ways in which $p$ nuclear
spins can coflip. The total spectral weight in a given polarization block is therefore
given by $f_p$, the probability that a particle will have polarization $p$. This
probability can be controlled by thermostatstical factors. For example, if we assume
that the nuclear spins are in equilibrium with the elctronic spin, then the
Boltzmann weights for $\bfI_i$ parallel and antiparallel to ${\bf s}_i$
are in the ratio $1\mathrel:\exp(-\be\om_n)$, where $\be = 1/k_B T$.
This leads to
\beq
f_p = (2\pi \sig_p^2)^{-1/2}e^{-(p - \bar p)^2/2\sig_p^2},
\label{fofp}
\eeq
where $\bar p = N_n\tanh(\be\om_n/2)$, and $\sig_p = N_n^{1/2}\sech(\be\om_n/2)$.
The dominant resonance line is thus that with zero polarization, and has a weight $f_0$.

Finally, we note that the effect of nuclear spins on magnetic particle MQC
is completely different from that of an oscillator bath.~\cite{chl,smr} While unlike
the latter, the
present problem does not seem to lend itself to the study of elegant mathematical
questions such as  renormalization group connections with the Kondo problem,~\cite{sc,bm}
it provides a novel and interestingly different example of how interaction with the
environment suppresess MQC. 

\section*{Acknowledgments}
This research is supported by the National Science Foundation via Grant number
DMR-9616749. I would also like to thank the members and the staff of the Institute
for Nuclear Theory, University of Washington, Seattle, for their warm hospitality
during the workshop on Tunneling in Complex Systems in the spring of 1997.
While the specific research reported herein was not carried out at the Institute,
my discussions with its members and various workshop participants have all
helped to increase my understanding of this subject.

\end{document}